\documentclass[preprint,prb]{revtex4-2}

\usepackage{graphicx}
\usepackage{dcolumn}
\usepackage{bm}
\usepackage{natbib}
\usepackage[english]{babel}
\usepackage{amsmath, amsfonts}
\usepackage{float}
\usepackage{xcolor}

\usepackage[caption=false]{subfig}

\begin{document}

\title{\boldmath An inversion problem for optical spectrum data via physics-guided machine learning. \unboldmath}

\author{Hwiwoo Park$^1$} \author{Jun H. Park$^{2, \dag}$} \author{Jungseek Hwang$^{1, *}$}

\affiliation{$^1$Department of Physics, Sungkyunkwan University, Suwon, Gyeonggi-do 16419, Republic of Korea \\ $^2$School of Mechanical Engineering, Sungkyunkwan University, Suwon, Gyeonggi-do 16419, Republic of Korea}
\date{\today}

\begin{abstract}

We propose the regularized recurrent inference machine (rRIM), a novel machine-learning approach to solve the challenging problem of deriving the pairing glue function from measured optical spectra. The rRIM incorporates physical principles into both training and inference and affords noise robustness, flexibility with out-of-distribution data, and reduced data requirements. It effectively obtains reliable pairing glue functions from experimental optical spectra and yields promising solutions for similar inverse problems of the Fredholm integral equation of the first kind.\\

Correspondence to [emails: $^{\dag}$jun.park@skku.edu and $^*$jungseek@skku.edu].

\end{abstract}


\maketitle

\section{Introduction}\label{introduction}

Experimental and theoretical investigations on high-temperature copper-oxide (cuprate) superconductors, since their discovery over 35 years ago \cite{bednorz:1986,wu:1987}, have afforded extensive results \cite{plakida:2010}. Despite these efforts, the microscopic electron-electron pairing mechanism for superconductivity remains elusive. In this regard, researchers have adopted innovative experimental techniques. Particularly, optical spectroscopy has the potential to elucidate the aforementioned pairing mechanisms because it is the only spectroscopic experimental method capable of providing quantitative physical quantities. The absolute pairing glue spectrum measured via optical spectroscopy may serve as a “smoking gun” evidence to address for this problem. The measured spectrum entails information concerning the pairing glue responsible for superconductivity. Extracting this glue function from the measured optical spectra via the decoding approach, which involves an inverse problem, contributes an essential aspect to the elucidation of high-temperature superconductivity.

The decoding-related inversion problem concerning physical systems is expressed as follows:
\begin{equation}\label{eq:generalized_allen}
\frac{1}{\tau^{\textrm{op}}(\omega,T)} = \int _0 ^{ \infty } d \Omega \,  I ^2 \chi ( \Omega, T) \, K( \omega, \Omega, T),
\end{equation}
which is referred to as the generalized Allen formula \cite{allen:1971,Shulga1991,Schachinger2006,Hwang2007}. Here $1/\tau^{\mathrm{op}}(\omega)$ is the optical scattering rate, $K(\omega, \Omega)$ is the kernel, and $I^2\chi(\omega)$ is the pairing glue function, which describes interacting electrons by exchanging the force-mediating boson. Further, $\chi(\omega)$ is the boson spectrum and $I$ denotes the electron–boson coupling constant. The kernel is given as
\begin{equation}\begin{aligned}\label{eq:shulga_kernel}
K( \omega, \Omega, T) = &\frac{ \pi }{ \omega } \left[ 2 \omega \coth \left( \frac{ \Omega }{ 2T } \right) - ( \omega + \Omega ) \coth \left( \frac{ \omega + \Omega }{ 2T } \right) \right.\\
&\left.  + ( \omega - \Omega ) \coth \left( \frac{ \omega - \Omega }{ 2T } \right) \right].
\end{aligned}\end{equation}
This is referred to as the Shulga kernel \cite{Shulga1991}. The goal was to infer the glue function from the optical scattering rate obtained using optical spectroscopy \cite{Hwang2007}.  Eq.~\eqref{eq:generalized_allen} can be expressed in a more general form as follows:
\begin{equation}\label{eq:fredholm_1st}
y(t) = \int _a ^b d\tau \, x(\tau) \, k(t, \tau),
\end{equation}
which is the Fredholm integral equation of the first kind. Here, $ k(t, \tau) $ is referred to as the kernel and determined by the underlying physics of the given problem, and $x(\tau)$ and $y(t)$ are physical quantities related to each other through the integral equation. Such inverse problems occur in many areas of physics \cite{Schachinger2006, Wazwaz2011, Yoon2018} and are known to be ill-posed \cite{Vapnik98}. The ill-posed nature arises from the instability of solutions in Eq. (3), where small changes in $y$ can lead to significant changes in $x$. Consequently, obtaining a solution to the inverse problem becomes challenging, particularly when observations are corrupted by noise.

Conventional approaches for solving inverse problems, expressed in the form of Eq.~\eqref{eq:fredholm_1st} include singular value decomposition (SVD) \cite{dordevic:2005}, least squares fit \cite{Hwang2006,heumen:2009}, maximum entropy method (MEM)~\cite{Schachinger2006, Hwang2007}, and Tikhonov regularization~\cite{Ito2014}. In particular, the MEM can effectively capture the key features of $x$, whereas the SVD and least-squares fit approaches, respectively, yield non-physical outputs and requiring a priori assumptions on the shape or size of $x$. At high noise levels, the MEM struggles to determine a unique amplitude ~\cite{Hwang2016} or or capture the peaks of $ x$~\cite{Fournier2020}. Despite its theoretical advantages in terms of convergence properties and well-defined solutions, Tikhonov regularization poses challenges with regard to the selection of appropriate regularization parameters ~\cite{Calvetti2000, Reichel2007,De2022}. 

In recent years, machine learning approaches have frequently been applied to inverse problems of the Fredholm integral of the first kind ~\cite{Fournier2020, Arsenault2017, Park2021}. Early approaches primarily utilized supervised learning~\cite{Yoon2018, Fournier2020, Park2021} demonstrating that machine learning results were comparable or superior to those of the conventional MEM and more robust against observation noise. Regardless of their successes, they are principally model-agnostic, resembling black box models wherein the physical model is solely used for generating training data but not leveraged during the inference procedure. Consequently, these methods lack explainability or reliability in their outputs and require substantial training data to achieve competitive results. This typically does not pose any difficulty in conventional machine learning domains, such as image classification, speech recognition, and text generation, wherein vast datasets are readily available, and explainability and reliability may be relatively less critical. In contrast, in scientific domains, a sound theoretical foundation must be provided for the output of the model, and data acquisition must be cost-effective. This is particularly crucial, because the data collection often necessitates expensive experiments or extensive simulations.

Various approaches have been explored to incorporate physics in the machine learning process. Projection is applied to the outcomes of supervised learning; this imposes physical constraints on $x$ such as normality, positivity, and compliance with the first and second moments \cite{Arsenault2017}. In a prior study \cite{Nguyen2021}, the forward model (Eq.~\eqref{eq:generalized_allen}) is added to the loss term for the training and serves as a regularizer. By utilizing the incorporated physics, these approaches can yield comparable results with significantly smaller training datasets as compared to conventional approaches. However, in these approaches, the application of physical constraints occurs at the last stage of the learning process. Consequently, they may not be sufficiently flexible to handle situations wherein the input significantly deviates from the training data. Adaptability to unseen data is crucial because the generated training data may not encompass all possible solutions, potentially leading to a bias in the output toward the training data.

To address this issue, we adopted a recurrent inference machine (RIM)~\cite{Putzky2017}. Compared to other supervised approaches, RIM exhibits exceptional capability to incorporate physical constraints throughout both the learning and inference processes through an iterative process. This implies that physical principles are not applied solely at the last stage of the learning process but are integrated throughout the learning and inference stages; thus, RIM has remarkable flexibility for solving complex inverse problems. Additionally, the RIM framework automates many hand-tuned optimization operations, streamlining the training procedure and achieving improved results \cite{Putzky2017, Andrychowicz2016}. The RIM framework has been applied to various types of image reconstruction ranging from tomographic projections~\cite{Adler2017} to astrophysics~\cite{Morningstar2019}. However, conventional RIM requires modification to achieve competitive results in case of ill-posed inverse problems such as the Fredholm integral equation of the first kind. Accordingly, we developed the regularized RIM (rRIM). Furthermore, we demonstrated that the rRIM framework is equivalent to iterative Tikhonov regularization ~\cite{Landweber1951, Yuan2019, Neumaier1998}. 

The proposed physics-guided approach significantly reduces the amount of training data required, typically by several orders of magnitude. To illustrate this aspect, we quantitatively compared the average test error losses of the rRIM with those of two widely used supervised learning approaches––a fully connected network (FCN) and convolutional neural network (CNN). Owing to the sound theoretical basis of the rRIM, it can adequately explain its outputs; this is a crucial feature in scientific applications and in contrast with purely data-driven black box models. We demonstrated the robustness of the rRIM in handling noisy data by comparing its performance with those of the FCN and CNN models which are well-known for their strong noise robustness compared to MEM, especially under high observational noise conditions~\cite{Yoon2018, Fournier2020}. The rRIM exhibited much better robustness than FCN and CNN for a wide range of noise levels. The rRIM exhibits superior flexibility in handling out-of-distribution (OOD) data than do the FCN and CNN approaches. The OOD data refers to data that significantly differs in characteristics, such as shape and distribution, from any of the training data. Finally, we applied the rRIM to the experimental optical spectra of optimally and overdoped Bi$_2$Sr$_2$CaCu$_2$O$_{8+\delta}$ (Bi-2122) samples. The results were shown to be comparable with those obtained using the widely employed MEM. As a result, rRIM can be a suggestive inversion method to analyze data with significant noise.

\section{Methods}\label{methods}

The generalized Allen formula, Eq.~\eqref{eq:generalized_allen}, can be written as
\begin{equation*}
1/ \tau ^{ \textrm{op}}( \omega, T ) = F( I ^2 \chi ( \omega, T )),
\end{equation*}
where $F( \cdot)$ represents an integral (forward) operator. Because of the linearity of an integral operator, this equation can be discretized into the following matrix equation:
\begin{equation}\label{eq:discretized_allen}
y = A x.
\end{equation}
For notational simplicity, we represent the glue function and optical scattering rate as $ x \in \mathbb{R}^n$ and $y \in \mathbb{R}^m $, respectively. $A$ is an $m \times n$ matrix containing the kernel information in Eq.~\eqref{eq:shulga_kernel}. Considering the noise in the experiment, Eq.~\eqref{eq:discretized_allen} can be written as
\begin{equation}\label{eq:discretized_allen_noise}
y = Ax + \eta,
\end{equation}
where \(\eta\) follows a normal distribution with zero mean and $ \sigma ^2 $ variance, that is, \(\eta \sim N(0,\sigma ^2 I)\). As the matrix $ A \in \mathbb{R}^{ m \times n } $ is often under-determined $ (m < n) $ and/or ill-conditioned (i.e., many of its singular values are close to $0$), inferring $x$ from $y$ requires additional assumption on the structure of $x$. On this basis, the goal of the inverse problem is formulated as follows:
\begin{equation}\label{eq:discretized_argmin}
\hat{x} = \arg \min_{x \in \mathcal{X}}  || y - Ax || ^2.
\end{equation}
Notably, the solution to Eq.~\eqref{eq:discretized_argmin} must satisfy two constraints: it minimizes the norm $|| \cdot ||$ and belongs to the data space $\mathcal{X}$, which represents the space of proper glue functions. The norm can be chosen based on the context of the problem; in this study, we selected $l_2$ norm.

The inverse problem expressed by Eq.~\eqref{eq:discretized_argmin}, with the prior information on $x$, can be considered a maximum a posteriori (MAP) estimation \cite{Figueiredo2007},
\begin{equation*}\label{eq:inverse_map}
\hat{x} = \arg \max_{x} \left\{ \log p(y|x) + \log p (x) \right\}.
\end{equation*}
The first term on the right-hand side represents the likelihood of the observation $y$ given $x$, which is determined using the noisy forward model (Eq.~\eqref{eq:discretized_allen_noise}). Further, the second term represents prior information regarding the solution space $\mathcal{X} $.

\begin{figure}[!htbp]
\includegraphics[width=0.90 \columnwidth]{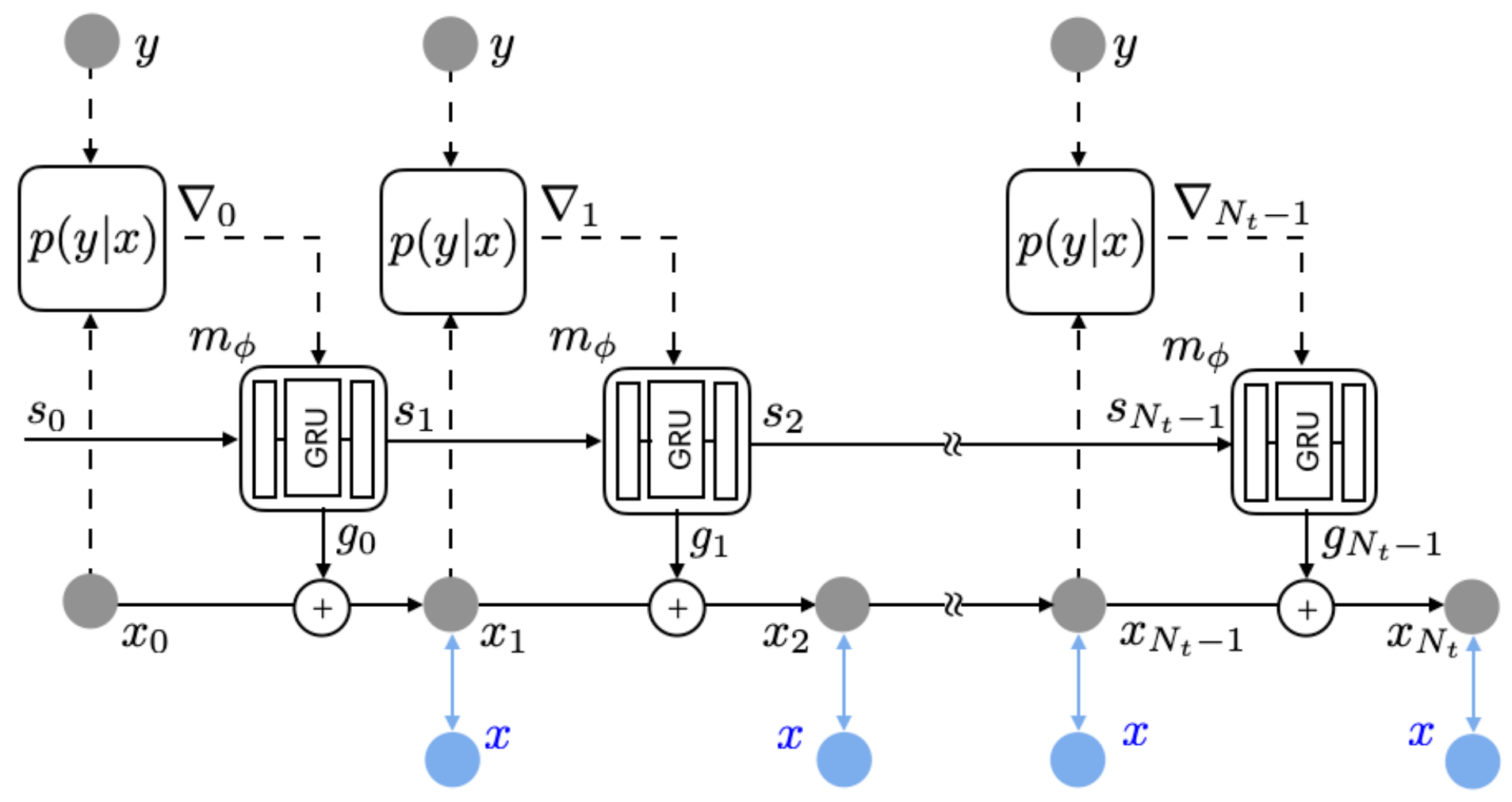}
\caption{(Color online) Schematic model of rRIM. Shown in blue is only used for training. Further details concerning the update network is provided in the Supplemental Material (adopted and modified from \cite{Putzky2017}).}
\label{fig:rim_rnn}
\end{figure}

The solution of the MAP estimation can be obtained using a gradient based recursive algorithm expressed as follows:
\begin{equation}\label{eq:inverse_gradient}
x _{ t + 1 } = x _t + \gamma  _t \nabla \, l(x _t ),
\end{equation}
where $ l (x) :=  \log p(y|x) + \log p (x) $ and $ \gamma _t $ is a learning rate. Determining the appropriate learning rate is challenging. In contrast, the RIM formulates Eq.~\eqref{eq:inverse_gradient} as
\begin{equation}\label{eq:inverse_rim}
x _{ t + 1 } = x _t + g _t (\nabla \, l(x _t ) ; \phi ),
\end{equation}
where $g _t $ is a deep neural network with learnable parameters $\phi$. More explicitly, RIM utilizes a recurrent neural network (RNN) structure as follows (Fig.~\ref{fig:rim_rnn})
\begin{equation}\label{eq:rim_rnn}\begin{aligned}
x _{ t + 1 } &= x _t + g _t ,\\
\left[
\begin{array}{c}
g _t  \\
s _{ t + 1 }
\end{array}
\right]
&= m _{ \phi } (\nabla _t, s _t ).
\end{aligned}\end{equation}
Given the observation $y$, the inference begins with a random initial value of $x _0 $. At each time step $t$, the gradient information $ \nabla _t := \nabla _x \log p(y|x) \vert _{x = x _t } $ is introduced into the update network $ m _{ \phi } $ along with a latent memory variable $ s _t $. The network's output  $ g _t $ is combined with the current prediction to yield the next prediction $ x _{ t + 1 } $. The memory variable $ s _{ t + 1 } $, another output of $ m _{ \phi } $, acts as a channel for the model to retain long term information, facilitating effective learning and inference during the iterative process \cite{Cho2014}.

In the training process, the loss is calculated by comparing the training data $x$ (shown in blue in Fig.~\ref{fig:rim_rnn}) with the model prediction $ x _t $ at each time step. The accumulated loss is calculated formulas follows:
\begin{equation}\label{eq:accumulated_error}
\mathcal{L} ( \phi ) = \sum _{ t = 1 } ^{N_t} w _t ( x _t ( \phi ) - x) ^2.
\end{equation}
Here, $ w _t $ represents a positive number (in this study, we set $ w _t = 1 $). The parameters $\phi$ indicate that $ x _t $ is obtained from the neural network $ m _{ \phi } $ and are updated using the backpropagation through time (BPTT) technique \cite{Goodfellow2016}. The solid edges in Fig.~\ref{fig:rim_rnn} illustrates the pathways for the propagation of the gradient $ \partial \mathcal{L}/\partial \phi $ to update $\phi$, whereas the dashed edges indicate the absence of gradient propagation. The trained model can perform an inference to estimate $x _t $ given the input $y$ without referencing the training data $x$.

In the conventional RIM, the gradient of the log-likelihood is given as
\begin{equation*}
\nabla _t = \frac{ 1 }{ \sigma ^2 } A ^T (y - A x _t ),
\end{equation*}
where, $A^{T}$ is the transpose of matrix $A$ (see the Supplemental Material for its derivation.) Owing to the inherent ill-posed nature of the aforementioned equation, it does not yield competitive results as compared to other machine learning approaches. To address this issue, we utilized the equivalence between the RIM framework and iterative Tikhonov regularization, as detailed in the Supplemental Material. Specifically, we utilized the following gradient derived from the preconditioned Landweber iteration~\cite{Neumaier1998} to formulate the rRIM algorithm:
\begin{equation}\label{eq:gradient}
\nabla _t = (A ^T A + h ^2 I) ^{ - 1 } A ^T (y - A x _t )
\end{equation}
Here, $h$ is a regularization parameter. Notably, the rRIM demonstrates unique flexibility in determining the appropriate regularization parameter, a task that is typically challenging.

A few important aspects are noteworthy. First, the noisy forward model in Eq.~\eqref{eq:discretized_allen_noise} plays a guiding role in both learning and inference throughout the iterative process, as shown in Eqs.~\eqref{eq:gradient}, which provides key physical insight into the model. Second, the gradient of log-prior information in Eq.~\eqref{eq:inverse_gradient} is implicitly acquired through the gradient of loss, $ \partial \mathcal{L}/\partial \phi $, which incorporates iterative comparisons between $ x _t $ and the training data $x$. Third, the learned optimizer (Eq.~\eqref{eq:inverse_rim}) yields significantly improved results compared to the vanilla gradient algorithm, Eq.~\eqref{eq:inverse_gradient} \cite{Andrychowicz2016}. Finally, the total number of time steps, $N_t$, serves as another regularization parameter that influences overfitting and underfitting. In the case of the rRIM, the output was robust to variations in this parameter. 

The iterative Tikhonov regularization algorithm functions as an optimization scheme, while the rRIM effectively addresses optimization problems using a recurrent neural network model. Specifically, rRIM optimizes the iterative Tikhonov regularization algorithm by replacing hand-designed update rules with learned ones, offering flexibility in selecting regularization parameters for iterative Tikhonov regularization. This enables automatic or lenient choices. As a result, the rRIM optimizes the iterative Tikhonov regularization and serves as an efficient and effective implementation of this method. As a result, the rRIM optimizes the iterative Tikhonov regularization and serves as an efficient and effective implementation of this method.

\section{Results and discussions}\label{results}

We generated a set of training data $ \left\{ x _n , y _n \right\} _{ n = 1 } ^N $, where $ x $ and $y$ represent $ I ^2 \chi $ and $ 1/ \tau ^{ \textrm{op}}$, respectively. Generating a robust training dataset that accurately represents the data space is challenging, especially in the context of inverse problems where the true characteristics of solutions are not readily available. In our study, based on existing experimental data, a parametric modele mploying Gaussian mixtures with up to $4$ Gaussians was used to create diverse $x$ values to simulate experimental results. Substituting the generated $x$ values into Eq.~\eqref{eq:discretized_allen} yields the corresponding $y$ values. The temperature was set at $100$ K. Datasets of varying sizes, $ N \in \left\{ 100, 1000, 10000,100000 \right\}$ were generated and split into training, validation, and test data sets with $ 0.8$, $ 0.1$, and $0.1$ ratios, respectively. Additionally, noisy samples were created by adding Gaussian noise with different standard deviations, $ \sigma  \in \left\{ 0.00001, 0.0001, 0.001, 0.01, 0.1 \right\} $. The noise amplitude added to each sample $y $ was determined by multiplying $\sigma$ with the maximum value of that sample. To ensure training stability, we scaled the data by dividing the $y$ values by $300$ (see the Supplemental Material for a more detailed description). It is worth noting that, depending on the characteristics of both the data and the model, a more diverse dataset could potentially span a broader range of the data space and uncover additional solutions.

We trained three models - FCN, CNN, and rRIM - on each dataset; the model architectures are presented in the Supplemental Material. All models were trained using the Adam optimizer~\cite{Kingma2015}. As regards the evaluation metrics, the mean-squared error (MSE) loss,
\[
\textrm{MSE} = \frac{ 1 }{ N } \sum _{ n = 1 } ^N  \left( x^{\textrm{training data}} _n  - x^{\textrm{model output}} _n  \right) ^2,
\]
was used for the FCN and CNN, and the accumulated error loss (Eq.~\eqref{eq:accumulated_error}) for the rRIM. Hyperparameter tuning was performed by using a validation set to obtain an optimal set of parameters for each model. Additionally, early stopping, as described in \cite{Goodfellow2016}, was applied to mitigate overfitting. For comparison, we calculated the same MSE loss using test datasets for all three models. Kernel smoothing was applied to reduce noise in the inference results. In the rRIM, we set $ N_t = 15 $ and $ h = 0.01$. For reference, we initially trained the FCN, CNN, and rRIM by using a noiseless dataset.

\begin{figure}[!htbp]
\subfloat{
{\includegraphics[width=0.90\columnwidth]{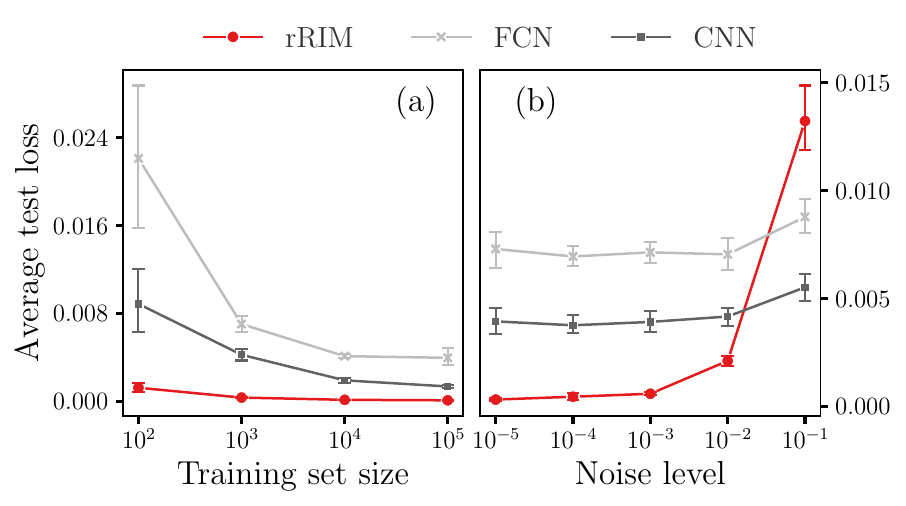}}
}
\\
\subfloat{
{\includegraphics[width=0.90\columnwidth]{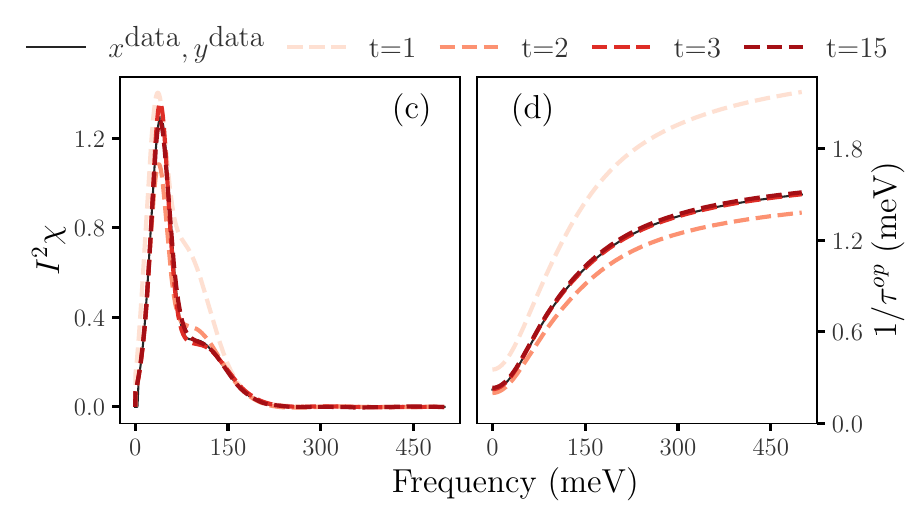}}
}
\caption{(Color online) Comparison of average test losses for rRIM, FCN, and CNN (a) for different training set sizes $N$ and (b) for different noise levels for $N$ = 1000. (c) Inference steps of rRIM for noiseless data for a selection of initial and final predictions (dotted lines) alongside with the true values of $x$ (solid line) and (d) their corresponding $y$ values, which are scaled down by 300.}
\label{fig:test_losses_pred_run}
\end{figure}

We conducted ten independent training runs for each model, each with different batch setups, and calculated the average test losses. Fig.~\ref{fig:test_losses_pred_run}(a) illustrates the trend of the average test losses for various training set sizes $N$. Across all three models, we observed a consistent pattern wherein the losses decreased as $N$ increased. Notably, the rRIM outperformed both the FCN and CNN across the entire range of $N$ in terms of error size and reliability. Additionally, our findings demonstrate that the CNN yields superior results compared to the FCN, as previously shown in \cite{Yoon2018}. 

Figs.~\ref{fig:test_losses_pred_run}(c) and (d) illustrate the inference process in the rRIM for the test data following training with the noiseless dataset when $ N = 1000 $. At each time step, the updated gradient information obtained from Eq.~\eqref{eq:gradient} is used to generate a new prediction. Fig.~\ref{fig:test_losses_pred_run}(c) shows the inference steps for $ t = 1, 2, 3,$ and $ 15\: (=N_t)$. The corresponding $y$ values are shown in Fig.~\ref{fig:test_losses_pred_run}(d); the intermediate results are omitted because of the absence of significant changes beyond a few initial time steps.

We evaluated the performance of the rRIM with noisy data by following a procedure similar to that employed for the noiseless case. For each of the three models, we conducted $10$ independent trials with varying noise levels. The average test losses for the $ N=1000$ data set are shown in Fig.~\ref{fig:test_losses_pred_run}(b). Notably, the rRIM exhibits significantly lower test losses than do the other models up to a certain noise threshold, beyond which its performance begins to deteriorate. This behavior can be attributed to the inherently ill-posed nature of the problem. In contrast to the model-agnostic nature of the FCN and CNN, the rRIM adopts an iterative approach that involves repeated application of the forward model. This iterative process can lead to error accumulation, influenced by factors such as the total number of time steps ($N_t$), the norm of $A$, and the noise intensity~\cite{Groetsch1993}. When the noise intensity is low, these factors may have a minimal impact on the overall error. However, as the noise intensity increases, their influence becomes more pronounced, potentially resulting in significant error accumulation. Similar patterns are observed for different training set sizes, as detailed in the Supplemental Material. Although it does not accommodate extremely high noise levels, the rRIM is suitable for a wide range of practical applications with moderate noise levels. It is worth noting that the superiority of FCN and CNN in noise robustness over MEM, as demonstrated in previous studies~\cite{Yoon2018, Fournier2020}, was observed at much lower noise levels than in the present study.

\begin{figure}[!htbp]
\subfloat{
\includegraphics[width=0.8\columnwidth]{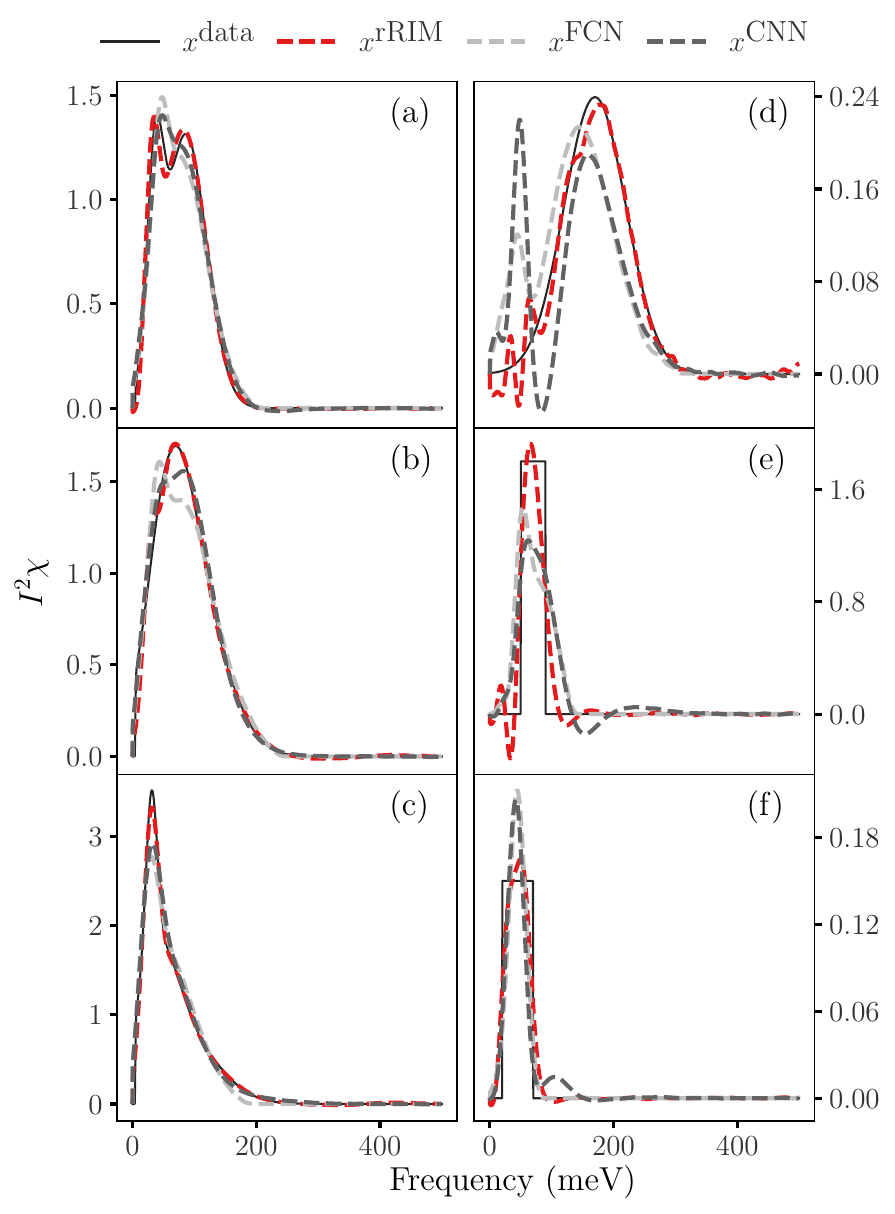}
}
\caption{(Color online) Comparison of prediction capabilities of rRIM, FCN, and CNN; for noisy test data samples: (a), (b), and (c); for OOD data samples: (d), (e), and (f).}
\label{fig:x_pred_samples}
\end{figure}

We compared the inference results of the rRIM with those of the FCN and CNN in Figs.~\ref{fig:x_pred_samples}(a), (b), and (c) by using the $ N = 1000$ with $ \sigma = 10^{-3} $ dataset. Inferences were made on the test data samples that were not part of the training process. Evidently, the rRIM accurately captures the true data height, whereas the FCN and CNN models tend to overshoot (a) and undershoot (c) it. Only the rRIM accurately replicated the shape of the peak (Fig.~\ref{fig:x_pred_samples}(b)). 

The ability of an algorithm to handle OOD data is crucial for its credibility, particularly in real-world scenarios where the solutions often lie beyond the scope of the dataset. To demonstrate the flexibility of the rRIM, we generated three sample datasets that differed significantly from the training datasets. The first dataset is a simple Gaussian distribution with a large variance, whereas the other two are square waves with different heights and widths (solid lines in Figs.~\ref{fig:x_pred_samples}(d), (e), and (f)). We inferred these data by using the rRIM, FCN, and CNN (the results are shown using dashed lines). These models were trained using a noiseless dataset of size $ N = 100000$. A Comparison of the results with the FCN and CNN models clearly indicates that the rRIM effectively captures the location, height, and width of the peaks in the provided data. Similar patterns are observed for trapezoidal and triangular waves, with detailed results provided in the Supplemental Material.

\begin{figure}[!htbp]
\begin{center}
\includegraphics[width=0.9\columnwidth]{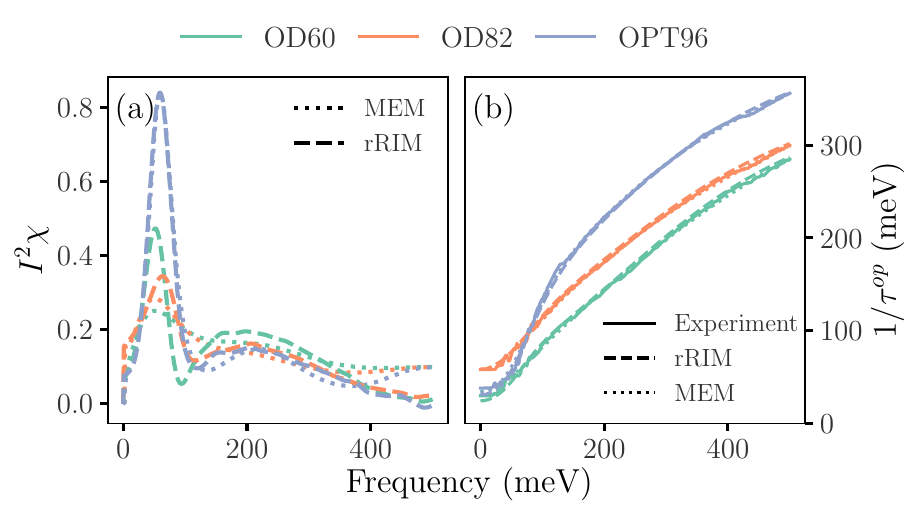}
\caption{(Color online) (a) Inference results of rRIM (dashed) with those of MEM (dotted) for experimental data. (b) Experimentally measured data (solid) with rRIM (dashed) and MEM (dotted) reconstruction. Samples (OD60, D82, and OPT96) are differentiated using the same color code in both figures. \label{fig:rim_vs_mem}}
\end{center}
\end{figure}

Finally, we applied the rRIM to real experimental data consisting of optically measured spectra, $1/\tau ^{\textrm{op}}(\omega)$ from one optimally doped sample ($T _c = 96$ K) and two overdoped samples ($T _c = 82 \textrm{ and } 60$ K) of Bi-2212, denoted as OPT96, OD82, and OD60, respectively (indicated using different colors in Fig.~\ref{fig:rim_vs_mem}). Here $T_c$ is the superconducting critical temperature. These measurements were performed at $T= 100$ K \cite{Hwang2007}. These experimental spectra were fed into the rRIM for inference. Fig.~\ref{fig:rim_vs_mem}(a) presents a comparison of the results of the rRIM (dashed line) with previously reported MEM results (dotted line) \cite{Hwang2007}. Fig.~\ref{fig:rim_vs_mem}(b) shows a comparison of the reconstructions of the optical spectra from the results of the rRIM (dashed line) and MEM (dotted line) by using Eq.~\eqref{eq:discretized_allen} with the experimental results (solid line). The rRIM results were competable to those of MEM.    

\section{Conclusions}\label{conclusion}

In this study, we devised the rRIM framework and demonstrated its efficacy in solving inverse problems involving the Fredholm integral of the first kind. By leveraging the forward model, we achieved superior results compared to pure supervised learning with significantly smaller training dataset sizes and reasonable noise levels. The rRIM shows impressive flexibility when handling OOD data. Additionally, we showed that the rRIM results were comparable to those obtained using MEM. Remarkably, we established that the rRIM can be interpreted as an iterative Tikhonov regularization procedure known as the preconditioned Landweber method \cite{Landweber1951, Yuan2019}. This characteristic indicates that the rRIM is an interpretable and reliable approach, making it suitable for addressing scientific problems.

Although our approach outperforms other supervised learning-based methods in many respects, it has certain limitations that warrant further investigation. First, we fixed the temperature in the kernel, limiting the applicability of our approach to experimental results at the same temperature. Extending this method for applicability in a wide range of temperatures would require a suitable training set. Another challenge concerns the generation of a robust training dataset that accurately reflects the data space required for a specific problem. This challenge is common to any machine learning approach to inverse problems that entails training data generation using the given forward model. Although the rRIM can partially address this issue by incorporating prior information through iterative comparison with the training data, more robust and innovative solutions are required. Accordingly, we posit that deep generative models \cite{Whang2020, Ongie2020} must be explored in this regard. While we have shown rRIM’s ability to handle OOD data, a comprehensive quantification of this capability has not been included in the current study. Acknowledging the significance of rRIM’s ability to manage OOD data, we intend to delve deeper into this aspect in future research endeavors. Finally, uncertainty evaluation must be considered for assessing the reliability of the output; this aspect was not addressed in the proposed approach.

%
%
\bibliographystyle{naturemag}
\bibliography{inverse_rim.bib}

\begin{thebibliography}{10}
\expandafter\ifx\csname url\endcsname\relax
  \def\url#1{\texttt{#1}}\fi
\expandafter\ifx\csname urlprefix\endcsname\relax\def\urlprefix{URL }\fi
\providecommand{\bibinfo}[2]{#2}
\providecommand{\eprint}[2][]{\url{#2}}

\bibitem{bednorz:1986}
\bibinfo{author}{Bednorz, J.~G.} \& \bibinfo{author}{Muller, A.}
\newblock \emph{\bibinfo{journal}{Z. Phys. B}} \textbf{\bibinfo{volume}{64}},
  \bibinfo{pages}{189} (\bibinfo{year}{1986}).

\bibitem{wu:1987}
\bibinfo{author}{Wu, M.~K.} \emph{et~al.}
\newblock \bibinfo{title}{Superconductivity at 93 k in a new mixed-phase
  \mbox{Y-Ba-Cu-O} compound system at ambient pressure}.
\newblock \emph{\bibinfo{journal}{Phys. Rev. Lett.}}
  \textbf{\bibinfo{volume}{58}}, \bibinfo{pages}{908} (\bibinfo{year}{1987}).

\bibitem{plakida:2010}
\bibinfo{author}{Plakida, N.}
\newblock \emph{\bibinfo{title}{High-temperature cuprate superconductors}}
  (\bibinfo{publisher}{Springer-Verlag GmbH Berlin, Heidelberg},
  \bibinfo{year}{2010}).

\bibitem{allen:1971}
\bibinfo{author}{Allen, P.~B.}
\newblock \bibinfo{title}{Electron-phonon effects in the infrared properties of
  metals}.
\newblock \emph{\bibinfo{journal}{Phys. Rev. B}} \textbf{\bibinfo{volume}{3}},
  \bibinfo{pages}{305} (\bibinfo{year}{1971}).

\bibitem{Shulga1991}
\bibinfo{author}{Shulga, S.~V.}, \bibinfo{author}{Dolgov, O.~V.} \&
  \bibinfo{author}{Maksimov, E.~G.}
\newblock \bibinfo{title}{Electronic states and optical spectra of htsc with
  electron-phonon coupling}.
\newblock \emph{\bibinfo{journal}{Phys. C Supercond. its Appl.}}
  \textbf{\bibinfo{volume}{178}}, \bibinfo{pages}{266} (\bibinfo{year}{1991}).

\bibitem{Schachinger2006}
\bibinfo{author}{Schachinger, E.}, \bibinfo{author}{Neuber, D.} \&
  \bibinfo{author}{Carbotte, J.~P.}
\newblock \bibinfo{title}{Inversion techniques for optical conductivity data}.
\newblock \emph{\bibinfo{journal}{Phys. Rev. B}} \textbf{\bibinfo{volume}{73}},
  \bibinfo{pages}{184507} (\bibinfo{year}{2006}).

\bibitem{Hwang2007}
\bibinfo{author}{Hwang, J.}, \bibinfo{author}{Timusk, T.},
  \bibinfo{author}{Schachinger, E.} \& \bibinfo{author}{Carbotte, J.~P.}
\newblock \bibinfo{title}{Evolution of the bosonic spectral density of the
  high-temperature superconductor \mbox{Bi$_2$Sr$_2$CaCu$_2$O$_{8+\delta}$}}.
\newblock \emph{\bibinfo{journal}{Phys. Rev. B}} \textbf{\bibinfo{volume}{75}},
  \bibinfo{pages}{144508} (\bibinfo{year}{2007}).

\bibitem{Wazwaz2011}
\bibinfo{author}{Wazwaz, A.-M.}
\newblock \bibinfo{title}{The regularization method for fredholm integral
  equations of the first kind}.
\newblock \emph{\bibinfo{journal}{Computers and Mathematics with Applications}}
  \textbf{\bibinfo{volume}{61}}, \bibinfo{pages}{2981--2986}
  (\bibinfo{year}{2011}).

\bibitem{Yoon2018}
\bibinfo{author}{Yoon, H.}, \bibinfo{author}{Sim, J.~H.} \&
  \bibinfo{author}{Han, M.~J.}
\newblock \bibinfo{title}{Analytic continuation via domain knowledge free
  machine learning}.
\newblock \emph{\bibinfo{journal}{Phys. Rev. B}} \textbf{\bibinfo{volume}{98}},
  \bibinfo{pages}{245101} (\bibinfo{year}{2018}).

\bibitem{Vapnik98}
\bibinfo{author}{Vapnik, V.~N.}
\newblock \emph{\bibinfo{title}{Statistical Learning Theory}}
  (\bibinfo{publisher}{John Wiley and Sons}, \bibinfo{year}{1998}).

\bibitem{dordevic:2005}
\bibinfo{author}{Dordevic, S.~V.} \emph{et~al.}
\newblock \bibinfo{title}{Extracting the electron-boson spectral function
  \mbox{$\alpha^2F(\omega)$} from infrared and photoemission data using inverse
  theory}.
\newblock \emph{\bibinfo{journal}{Phys. Rev. B}} \textbf{\bibinfo{volume}{71}},
  \bibinfo{pages}{104529} (\bibinfo{year}{2005}).

\bibitem{Hwang2006}
\bibinfo{author}{Hwang, J.} \emph{et~al.}
\newblock \bibinfo{title}{\mbox{$a$}-axis optical conductivity of detwinned
  ortho-ii \mbox{YBa$_2$Cu$_3$O$_{6.50}$}}.
\newblock \emph{\bibinfo{journal}{Phys. Rev. B}} \textbf{\bibinfo{volume}{73}},
  \bibinfo{pages}{014508} (\bibinfo{year}{2006}).

\bibitem{heumen:2009}
\bibinfo{author}{van Heumen, E.} \emph{et~al.}
\newblock \bibinfo{title}{Optical determination of the relation between the
  electron-boson coupling function and the critical temperature in
  high-\mbox{$T_c$} cuprates}.
\newblock \emph{\bibinfo{journal}{Phys. Rev. B}} \textbf{\bibinfo{volume}{79}},
  \bibinfo{pages}{184512} (\bibinfo{year}{2009}).

\bibitem{Ito2014}
\bibinfo{author}{Ito, K.} \& \bibinfo{author}{Jin, B.}
\newblock \emph{\bibinfo{title}{Inverse problems: Tikhonov theory and
  algorithms}}, vol.~\bibinfo{volume}{22} (\bibinfo{publisher}{World
  Scientific}, \bibinfo{year}{2014}).

\bibitem{Hwang2016}
\bibinfo{author}{Hwang, J.}
\newblock \bibinfo{title}{{Intrinsic temperature-dependent evolutions in the
  electron-boson spectral density obtained from optical data}}.
\newblock \emph{\bibinfo{journal}{Sci. Rep.}} \textbf{\bibinfo{volume}{6}},
  \bibinfo{pages}{23647} (\bibinfo{year}{2016}).

\bibitem{Fournier2020}
\bibinfo{author}{Fournier, R.}, \bibinfo{author}{Wang, L.},
  \bibinfo{author}{Yazyev, O.~V.} \& \bibinfo{author}{Wu, Q.~S.}
\newblock \bibinfo{title}{Artificial neural network approach to the analytic
  continuation problem}.
\newblock \emph{\bibinfo{journal}{Phys. Rev. Lett.}}
  \textbf{\bibinfo{volume}{124}}, \bibinfo{pages}{056401}
  (\bibinfo{year}{2020}).

\bibitem{Calvetti2000}
\bibinfo{author}{Calvetti, D.}, \bibinfo{author}{Morigi, S.},
  \bibinfo{author}{Reichel, L.} \& \bibinfo{author}{Sgallari, F.}
\newblock \bibinfo{title}{Tikhonov regularization and the l-curve for large
  discrete ill-posed problems}.
\newblock \emph{\bibinfo{journal}{Journal of computational and applied
  mathematics}} \textbf{\bibinfo{volume}{123}}, \bibinfo{pages}{423}
  (\bibinfo{year}{2000}).

\bibitem{Reichel2007}
\bibinfo{author}{Reichel, L.}, \bibinfo{author}{Sadok, H.} \&
  \bibinfo{author}{Shyshkov, A.}
\newblock \bibinfo{title}{Greedy tikhonov regularization for large linear
  ill-posed problems}.
\newblock \emph{\bibinfo{journal}{International Journal of Computer
  Mathematics}} \textbf{\bibinfo{volume}{84}}, \bibinfo{pages}{1151}
  (\bibinfo{year}{2007}).

\bibitem{De2022}
\bibinfo{author}{De~Vito, E.}, \bibinfo{author}{Fornasier, M.} \&
  \bibinfo{author}{Naumova, V.}
\newblock \bibinfo{title}{A machine learning approach to optimal tikhonov
  regularization i: affine manifolds}.
\newblock \emph{\bibinfo{journal}{Analysis and Applications}}
  \textbf{\bibinfo{volume}{20}}, \bibinfo{pages}{353} (\bibinfo{year}{2022}).

\bibitem{Arsenault2017}
\bibinfo{author}{Arsenault, L.-F.}, \bibinfo{author}{Neuberg, R.},
  \bibinfo{author}{Hannah, L.~A.} \& \bibinfo{author}{Millis, A.~J.}
\newblock \bibinfo{title}{Projected regression method for solving fredholm
  integral equations arising in the analytic continuation problem of quantum
  physics}.
\newblock \emph{\bibinfo{journal}{Inverse Problems}}
  \textbf{\bibinfo{volume}{33}}, \bibinfo{pages}{115007}
  (\bibinfo{year}{2017}).

\bibitem{Park2021}
\bibinfo{author}{Park, H.}, \bibinfo{author}{Park, J.~H.} \&
  \bibinfo{author}{Hwang, J.}
\newblock \bibinfo{title}{Electron-boson spectral density functions of cuprates
  obtained from optical spectra via machine learning}.
\newblock \emph{\bibinfo{journal}{Phys. Rev. B}}
  \textbf{\bibinfo{volume}{104}}, \bibinfo{pages}{235154}
  (\bibinfo{year}{2021}).

\bibitem{Nguyen2021}
\bibinfo{author}{Nguyen, H.~V.} \& \bibinfo{author}{Bui-Thanh, T.}
\newblock \bibinfo{title}{Tnet: A model-constrained tikhonov network approach
  for inverse problems}.
\newblock \emph{\bibinfo{journal}{arXiv preprint arXiv:2105.12033}}
  (\bibinfo{year}{2021}).

\bibitem{Putzky2017}
\bibinfo{author}{Putzky, P.} \& \bibinfo{author}{Welling, M.}
\newblock \bibinfo{title}{{Recurrent inference machines for solving inverse
  problems}}.
\newblock \emph{\bibinfo{journal}{arXiv preprint arXiv:1706.04008}}
  (\bibinfo{year}{2017}).

\bibitem{Andrychowicz2016}
\bibinfo{author}{Andrychowicz, M.} \emph{et~al.}
\newblock \bibinfo{title}{Learning to learn by gradient descent by gradient
  descent}.
\newblock In \emph{\bibinfo{booktitle}{Proceedings of the 30th International
  Conference on Neural Information Processing Systems}}, NIPS'16,
  \bibinfo{pages}{3988} (\bibinfo{publisher}{Curran Associates Inc.},
  \bibinfo{address}{Red Hook, NY, USA}, \bibinfo{year}{2016}).

\bibitem{Adler2017}
\bibinfo{author}{Adler, J.} \& \bibinfo{author}{Oktem, O.}
\newblock \bibinfo{title}{Solving ill-posed inverse problems using iterative
  deep neural networks}.
\newblock \emph{\bibinfo{journal}{Inverse Probl.}}
  \textbf{\bibinfo{volume}{33}}, \bibinfo{pages}{124007}
  (\bibinfo{year}{2017}).

\bibitem{Morningstar2019}
\bibinfo{author}{Morningstar, W.~R.} \emph{et~al.}
\newblock \bibinfo{title}{Data-driven reconstruction of gravitationally lensed
  galaxies using recurrent inference machines}.
\newblock \emph{\bibinfo{journal}{The Astrophysical Journal}}
  \textbf{\bibinfo{volume}{883}}, \bibinfo{pages}{14} (\bibinfo{year}{2019}).

\bibitem{Landweber1951}
\bibinfo{author}{Landweber, L.}
\newblock \bibinfo{title}{An iteration formula for fredholm integral equations
  of the first kind}.
\newblock \emph{\bibinfo{journal}{American Journal of Mathematics}}
  \textbf{\bibinfo{volume}{73}}, \bibinfo{pages}{615} (\bibinfo{year}{1951}).

\bibitem{Yuan2019}
\bibinfo{author}{Yuan, D.} \& \bibinfo{author}{Zhang, X.}
\newblock \bibinfo{title}{An overview of numerical methods for the first kind
  fredholm integral equation}.
\newblock \emph{\bibinfo{journal}{SN Applied Sciences}}
  \textbf{\bibinfo{volume}{1}}, \bibinfo{pages}{1178} (\bibinfo{year}{2019}).

\bibitem{Neumaier1998}
\bibinfo{author}{Neumaier, A.}
\newblock \bibinfo{title}{Solving ill-conditioned and singular linear systems:
  A tutorial on regularization}.
\newblock \emph{\bibinfo{journal}{SIAM Review}} \textbf{\bibinfo{volume}{40}},
  \bibinfo{pages}{636} (\bibinfo{year}{1998}).

\bibitem{Figueiredo2007}
\bibinfo{author}{Figueiredo, M. A.~T.}, \bibinfo{author}{Nowak, R.~D.} \&
  \bibinfo{author}{Wright, S.~J.}
\newblock \bibinfo{title}{Gradient projection for sparse reconstruction:
  Application to compressed sensing and other inverse problems}.
\newblock \emph{\bibinfo{journal}{IEEE Journal of Selected Topics in Signal
  Processing}} \textbf{\bibinfo{volume}{1}}, \bibinfo{pages}{586}
  (\bibinfo{year}{2007}).

\bibitem{Cho2014}
\bibinfo{author}{Cho, K.} \emph{et~al.}
\newblock \bibinfo{title}{Learning phrase representations using {RNN}
  encoder-decoder for statistical machine translation}.
\newblock In \emph{\bibinfo{booktitle}{Proceedings of the 2014 Conference on
  Empirical Methods in Natural Language Processing ({EMNLP})}},
  \bibinfo{pages}{1724} (\bibinfo{publisher}{Association for Computational
  Linguistics}, \bibinfo{address}{Doha, Qatar}, \bibinfo{year}{2014}).

\bibitem{Goodfellow2016}
\bibinfo{author}{Goodfellow, I.}, \bibinfo{author}{Bengio, Y.} \&
  \bibinfo{author}{Courville, A.}
\newblock \emph{\bibinfo{title}{Deep Learning}} (\bibinfo{publisher}{MIT
  Press}, \bibinfo{year}{2016}).

\bibitem{Kingma2015}
\bibinfo{author}{Kingma, D.~P.} \& \bibinfo{author}{Ba, J.}
\newblock \bibinfo{title}{Adam: A method for stochastic optimization}.
\newblock In \bibinfo{editor}{Bengio, Y.} \& \bibinfo{editor}{LeCun, Y.} (eds.)
  \emph{\bibinfo{booktitle}{3rd International Conference on Learning
  Representations, {ICLR} 2015, San Diego, CA, USA, May 7-9, 2015, Conference
  Track Proceedings}} (\bibinfo{year}{2015}).

\bibitem{Groetsch1993}
\bibinfo{author}{Groetsch, C.~W.}
\newblock \emph{\bibinfo{title}{Inverse problems in the mathematical
  sciences}}, vol.~\bibinfo{volume}{52} (\bibinfo{publisher}{Springer},
  \bibinfo{year}{1993}).

\bibitem{Whang2020}
\bibinfo{author}{Whang, J.}, \bibinfo{author}{Lei, Q.} \&
  \bibinfo{author}{Dimakis, A.~G.}
\newblock \bibinfo{title}{{Compressed Sensing with Invertible Generative Models
  and Dependent Noise}}.
\newblock \emph{\bibinfo{journal}{CoRR}}
  \textbf{\bibinfo{volume}{abs/2003.08089}} (\bibinfo{year}{2020}).

\bibitem{Ongie2020}
\bibinfo{author}{Ongie, G.} \emph{et~al.}
\newblock \bibinfo{title}{Deep learning techniques for inverse problems in
  imaging}.
\newblock \emph{\bibinfo{journal}{IEEE J. Sel. Areas Inf. Theory}}
  \textbf{\bibinfo{volume}{1}}, \bibinfo{pages}{39} (\bibinfo{year}{2020}).

\end{thebibliography}

\newpage

\newpage

\newpage

\newpage

\newpage

\noindent {\bf Author Contribution} JH and JP wrote the main manuscript. HP and JP performed the calculations for getting the data in the paper. All authors reviewed the manuscript.
\\ \\

\noindent {\bf Acknowledgements} This paper was supported by the National Research Foundation of Korea (NRFK Grants No. 2017R1A2B4007387 and No. 2021R1A2C101109811).
\\ \\

\noindent {\bf Competing Interests} The authors declare that they have no competing financial interests.
\\ \\

\noindent {\bf Availability of Data and Materials} The datasets used and/or analysed during the current study available from the corresponding author on reasonable request.
\\  \\

\noindent {\bf Correspondence} Correspondence and requests for materials should be addressed to Jungseek Hwang (email: jungseek@skku.edu) and Jun H. Park (jun.park@skku.edu).

\end{document}